# Study of the fundamental physical principles in atmospheric modeling based on identification of atmosphere - climate control factors

Version 3.0 (July 2010)

PART 2

GAIA PARADIGM

A BIOTIC ORIGIN OF THE POLAR SUNRISE

ARCTIC BROMINE EXPLOSION

## TABLE OF CONTENTS

# ABSTRACT
**Index: Springtime Arctic bromine explosion, ocean bromine concentrating vs. nitrogen fixation by soil bacteria, Gaia paradigm, biogeochemical cycling, multiple unity and interminability of the Earth's operations**

The tropospheric composition in the springtime Arctic is influenced by the natural surface emissions of bromine species. The associated phenomena of the sunrise Arctic bromine explosion include ozone depletion events (ODEs) in the atmospheric boundary layer (ABL) at high latitudes and the consequent perturbation of the tropospheric ozone field downward to mid latitudes. Bromine explosion appears to be an eminent stage in the biogeochemical cycles of oxygen and bromine. During the springtime transition, bromine explosion determines the Arctic and Northern Hemisphere's (NH) climate stability and changes. The main attention is given to discussion of the natural causes and regularities of bromine explosion. In order to emphasize the conceptual aspects of phenomena, we make a notice of marine microbiota as a prime driving factor for the elemental biogeochemical cycles of oxygen and bromine.

The resemblance is drawn between biotic processes responsible for the similar life-sustaining functions. One of the important conclusions of this study is on similarity between bromine concentrating as by-product of the marine microbial activities and nitrogen fixation by soil bacteria. In both cases, microbial organisms and their food webs have maintained life-sustaining superficial environments throughout geological ages of the Earth.

Following to von Neumann's opportunistic belief ("The Mathematician") that science is successful when it summarizes facts into a few fundamental principles, we use the phenomena of the Arctic bromine explosion to render the identification of the planet's atmosphere and climate controlling factors.

## § 1 INTERPLAY OF OXYGEN AND BROMINE BIOGEOCHEMISTRY
**Index: Phenomenology of Arctic bromine explosion, interception of oxygen and bromine biogeochemical cycles, ozone-oxygen conversion, critical role of free oxygen**

In February-March, at sunlit, erosion and thawing of the Arctic land and subsea permafrost expose microbial biota to the solar radiation. The erosion of permafrost in sunlit conditions leads to restoration of the microbial metabolism and decomposition of by-product bromine salt aggregates. The most possible locations for the commencement of bromine explosion phenomenon are coastal sites with the intensive erosion rates under mechanical waves, chemical and salt weathering processes. Shoreline erosion is one of the most rapid and observable. The coastal sensitivity is known for many thousands kilometers of the Arctic coast line. Early episodes of ABL bromine explosion start at the different shoreline locations and characterize by the high bromine content and complete tropospheric ozone depletion up to altitudes of 500 - 1500 m.

Phenomenology of bromine explosion includes four sequential processes:

1) Decomposition of by-products of marine microbiota activities introduces catalytic bromine compounds to the Arctic environment
2) Bromine compounds are interposed to the Arctic sea ice/ snow pack and atmosphere
3) The blowing snow over the Arctic surface distributes bromine pollutants all over the sea ice
4) In ABL, reactive bromine catalytically depletes surface and near-surface ozone

Phenomenology shows the exact synchronization of the interrelated biotic-abiotic processes.

In remote Arctic region, by altering local surface ozone-oxygen concentrations, the biotic bromine explosion controls over the ABL chemistry. With a cold polar air masses transfer, the perturbations on the Arctic atmospheric composition spread downward to mid latitudes. We provide an example of the qualitative analysis of ozone- oxygen conversion, illustrated by GEM model results.

Both outgassing to atmosphere and biotic activities are integrated into the biogeochemical cycling of chemical elements. Biogeochemical cycling carries out exchange of chemical elements and information between the biotic and abiotic processes and strengthens the life favorable environment. Synchronization of biotic-abiotic processes concerts the temporal development of the Earth's atmosphere-climate system at global and regional scales. In our research, we focus on the interactions and the synchronization of the biogeochemical cycles of oxygen and bromine species taking place in the present marine boundary layer (MBL) of the Arctic troposphere. Herein, the related phenomenology and a paired biotic-abiotic approach are offered upon the Gaia paradigm. We dwell only on the solar aspect of synchronization and yield on the foundational principle of interminable multiple unity of the Earth's operations.

Biotic and abiotic processes perplex the utilization of bromine explosion. Biotic processes include activities of the Arctic ocean marine community. Stabilization of dissolved oxygen and normalization of surface ozone are crucially important for the marine life. Following the polar sunrise, the productive season of the Arctic ocean is starting in months of March-April. Sometime after the Julian day 130, the net marine community production reaches its maximum. Throughout the productive season, the Arctic ocean is a source of the atmospheric oxygen. However, for the marine production to approach a peak, life favorable conditions should be established.

Spontaneous bromine explosion happens at polar sunrise. Polar sunrise is characterized by the excess of ozone flow from stratosphere to ABL. In late winter and early spring, Arctic surface waters serve as an atmospheric oxygen sink. The rates of biogeochemical exchange, including oxygen fluxes, are higher at coastal permafrost and shallow shelf sites. The initial phase of bromine explosion is associated with the coastal locations. We know that these sites are particularly important for biota activities listed under the microbial processes.

Ozone- oxygen transformation is a major outcome of the bromine catalytic cycle. Prime source of ABL ozone in Arctic is the Northern Hemisphere stratospheric ozone.

During the bromine explosion season, ABL ozone is effectively converted to molecular oxygen. Enriched by bromine species, merely ozone-poor, the air masses transport portions of by-conversion oxygen all over Arctic and toward mid latitudes. Oxygen also sinks into the Arctic surface water column and feeds to oxygen maximum zones of the Arctic ocean. While the total mass of oxygen in atmosphere remains at constant levels, the tropospheric ozone-oxygen conversion is resulted in the intensification of oxygen fluxes. Intensification of oxygen fluxes powers the non equilibrium Earth's system and recreates its stability.

By a complex of biotic-abiotic processes, bromine of the Arctic origin is redistributed over NH. Some of it is deposited, also in the bromine-poor environments. Through the food web connections, redistribution favors pool of the Earth's life forms which metabolism includes the bromide ions.

Stability and development of Earth's system guarantee stability and development of Earth's biota. Earth's biotic-abiotic operations are linked to each other through the biogeochemical cycling. Progress of "abiotic" (chemical) processes is managed by microbiota which frontally participates in the biogeochemical cycles ([21]).
Interception of bromine and oxygen cycles during the seasonal Arctic bromine explosion displays the extraordinary significance of Arctic paleo microbiota and bromine ocean emissions.

## § 2 OZONE-OXYGEN TRANSFORMATIONS IN THE ARCTIC ABL
**Index: Gaseous phase bromine catalytic cycle, dissolved ocean oxygen and oxygen photosynthetic microbiota, oxygen fluxes over the atmosphere-ocean barrier**

Complex phenomena of bromine activation at low temperatures in Arctic troposphere is available due to the sea ice/ snow pack barrier between Arctic atmospheric boundary layer and Arctic ocean surface mixed layer.
Inorganic bromine compounds Br, BrO, and HOBr are reactive and have the ability to destroy ozone ([1]). Gas phase catalytic destruction of ozone is present both in stratosphere and troposphere.
In the presence of frozen water content and high ozone concentrations, yet in the different atmospheric layers, reactive bromine regeneration is associated with a similar heterogeneous reaction mechanism at low temperatures**.**

Introduced to the Canadian Global Environmental Model (GEM), bromine chemistry is presented in [8]. Bromine compounds react with a number of tropospheric constituents. General bromine modeling includes photochemistry, gas phase reactions, heterogeneous chemistry, dry and wet deposition and transport processes. Responsible for ozone depletion, bromine chemical mechanism necessitates bromine species recycling on sea salt and sulfur aerosols. Bromine flux from sea ice/ snow pack reflects on the bromine inventories which are incorporated onto porous matrix of the Arctic ice pack.

## § 2.1. FATE OF MOLECULAR OXYGEN IN THE SUNRISE ARCTIC ABL
**Index: Significance of ozone-oxygen transformation for the Arctic marine life, metabolic activities and extreme climate conditions, maintenance of the viable dissolved oxygen levels by microbial organisms**

Ozone- oxygen conversion is a major outcome of the gas phase bromine catalytic cycle:

$2O_3 \rightarrow 3O_2$

(R1) $Br + O_3 \rightarrow BrO + O_2$

(R2) $BrO + BrO \rightarrow Br + Br + O_2$

(R3) $BrO + BrO \rightarrow Br_2 + O_2$

While the given above catalytic mechanism is usually discussed in regard to the chemical destruction of tropospheric ozone, we explore the fate of molecular oxygen in the sunrise MBL.

Free oxygen is critical to the life of Earth's organisms, and especially for the oxygen photosynthetic microbiota and its food webs. Abundance of free oxygen in atmosphere is a result of biogeochemical cycling. There is an evident link between the dissolved in ocean oxygen and the amounts of atmospheric oxygen on surface. Maintenance of the life favorable amounts of dissolved oxygen (DO) is performed by means of biogeochemical cycling of elements throughout all Earth's system's compartments, which involves earth atmosphere compartment. Biogeochemical cycling of oxygen includes interrelated cycles of free and combined oxygen. In this study, we consider only a cycle of free oxygen.

Marine organisms depend on the levels of dissolved oxygen. Since genesis, paleo marine microbiota has regulated the viable dissolved oxygen levels. Marine microbial life is primarily responsible for the maintaining of the current standard DO amounts. Paleo climate changes are often referred to the trends in the dissolved oxygen. The reciprocal connection between climate and dissolved oxygen changes is based on the optimal for biota (a) planetary and local energy balance and (b) planetary and local entropy production. Unity of Earth's system is built on the biotic-abiotic coupling, with microbiota as a leading force of the biogeochemical cycling. The interminable multiple unity of Earth's system's operations is a principle clause for the reasonable, intelligent, functioning of the Earth's system.

For the biodiversity of the modern large life forms, the fast oxygen cycle with its photosynthesis production of oxygen is exceptionally notable. Half of the oxygenic photosynthesis production is associated with the World Ocean community. The recent detection of viable ancient photosynthetic organisms from the Arctic (3 million years) and Antarctic (5 million years) permafrost bedrock represent an important achievement in cryobiology ([10] and [21]). The life cycle of Arctic marine microbiota consists of the separate conservation and production periods. At conservation periods, large amounts of microbiota are trapped in permafrost. At productive periods, microbial biota gets to the Arctic ocean, grows and reproduces.

During the geological periods, Arctic region experienced extremes of solar radiation and large fluctuations of the geomagnetic field. Arctic has a maritime climate. For the 200 million years from Jurassic to Holocene, Arctic and circum-arctic sedimentary records show components which indicate to adverse environmental development. Contemporary Arctic troposphere has its unique place within the Earth's atmosphere. Periods of darkness and daylight are prolonged for months. In winter and early spring, the temperatures are cold and very cold. Nighttime and springtime Arctic troposphere has a low (8-10 km) tropopause above and sea ice/ snow pack surface below. In winter and spring, atmospheric pressure gradients are strong. For instance, winter frequent cyclones are associated with the extension of the Icelandic Low northward and the Aleutian Low. High pressure areas are developed over Siberia, Beaufort- Chukchi Sea and Canadian Archipelago. Although the springtime atmospheric patterns turn in milder, the Arctic biota gigantic effort to sustain sets up in the extreme synoptic- climatic situation.

For geological times, ozone-oxygen transformation in the Arctic marine boundary layer has had a particular significance for the local marine life. As known, at the time of the bromine explosion season, atmospheric oxygen sinks into the surface water column. For subarctic locations, oxygen flux into ocean is estimated at 0.6- 0.3 mol $O_2$ m$^{-2}$ year$^{-1}$ ([23]). Temperature of the ocean surface mixed layer (SML) is extremely responsive to the surface conditions, including solar radiation, presence of sea ice/ snow pack and air temperature. Because the solubility of oxygen increases polarward (the SML temperature decreases), there is a large territory where the arctic oxygen fluxes should be higher than subarctic fluxes. There are no available reports on arctic oxygen fluxes or their average. However, indication of oxygen fluxes to SML is given by the dissolved oxygen amounts. The dissolved oxygen amounts of about 15mg/L are measured in the Arctic surface waters ([5]). It is 5 times more than the average dissolved amounts modeled without consideration of the biotic–abiotic interaction. Measured springtime DO levels are usually close to the saturated level.

## § 3 PHYSICAL AND BIOLOGICAL PRECURSORS OF PHENOMENA
**Index: Elevated amounts of surface ozone, sunlit and low temperature conditions, bromine pollution of Arctic sea ice/ snow pack and Arctic ABL, dissolved oxygen in SML of Arctic ocean**

In springtime, Arctic surface-atmosphere interface is represented by ocean covered by sea ice/ snow pack, open water (leads and polynias) and sea ice/ snow pack above land permafrost. Bromine explosion starts along the coastal locations. Arctic abrasive coasts and shelves contain a huge supply of viable marine microbial life forms. The special metabolism of paleo marine microorganisms guarantees their superb adaptability to the extreme climate conditions, such as lack of energy and nutrients.
Main planet's sources for bromine are seawater, lakes and well brines. They are serving a habitat for the marine biota. The original ancient marine biota is the microbiota. In Arctic, at cold temperatures, the viable microbiota is frozen in permafrost. Some permafrost repositories may be very old, but they are not dead.

Microbial activity is a major force behind the ocean bromine concentrating. Concentrating of bromine concentrating has a number of the possible pathways ([19]). For example, along the seashore, microbial population could coat bromine salt aggregates with a sort of vanish that prevent salt from dissolving back and running away from the shallow shelf region. Instantly, the paleo microbial organisms use brominated metabolites for the synthesis of essential amino acids. They also use bromine pigmentation to protect themselves and their spores against UV radiation. Bromine explosion starts with the returning of sunlit. The erosion of permafrost is the next to sunlit precursor to bromine injection to Arctic ABL. The bromine salt aggregates are carried into the snow pack/ sea ice at coastal locations and then redistributed by blowing snow and stormy winds over the Arctic. Long-distance pollution by blowing snow and stormy winds leads to the enlarged bromine emissions coverage. Most likely, the natural surface bromine emissions cover the most of the Arctic sea ice map. There is an obvious similarity between bromine concentrating by marine microbiota and nitrogen fixation by soil bacteria. The atmospheric concentration of nitrous oxide $N_2O$ of about 0.33 ppm is a chemical anomaly characteristic of the Earth and a consequence of its biosphere (P. Crutzen, SCOPE 21, ICSU). Like BrO, nitrous oxide $N_2O$ also reacts with ozone. Both release of $N_2O$ from soil and bromine explosion over the Arctic ocean modify ozone content in atmosphere. Similar to nitrogen fixation, the Arctic bromine concentrating and the other pathways of bromine concentrating by microbiota, attend on the same sensible metabolic functioning.

At first, bromine compounds are introduced to sea ice/ snow pack at habitat locations. Then the blowing snow distributes bromine pollutants over the Arctic sea ice coverage. In a process of snow-atmosphere exchange, bromine compounds are emitted to troposphere. Because Arctic winter tropopause is low, and stratospheric ozone gets to the marine boundary layer, dark time troposphere has high ozone content. In MBL, aircraft observations recorded ozone amounts up to 300 ppb. Based on the observation data and life-regulated connection between the ABL and tropopause ozone levels, the mean of surface ozone field is estimated as 50 ppb ([8]).

First episodes of bromine explosion show high vertical column (VC) BrO in the vicinity of coastal locations. Due to the atmospheric circulation and blowing snow pollution, bromine explosion/ ABL ozone depletion may be observed almost everywhere in a region. Bromine recycling on aerosols increases the life time of the phenomenon, and long-distant transport of bromine compounds out of the Arctic region becomes possible. As GOME data shows, in April and May, the bromine-rich air masses reach middle latitudes of the Northern Hemisphere and create the perturbation of ozone field. In the middle of summer, the phenomenon is ceasing. Since perturbations of the ozone field play a leading role in the variations of energy budgets, seasonal springtime perturbations are especially significant at solar minima.

Season of bromine explosion ends with normalization of surface ozone amounts and stabilization of dissolved oxygen in the Arctic waters. Stabilization of dissolved oxygen and normalization of surface ozone are resolute for the microbiota and for the all ocean food web. Involved in the biogeochemical cycling, the life–sustaining biotic activities clearly aim on (a) preserving the biodiversity and (b) balance with the environment. The global impact of Arctic bromine explosion/ ABL ozone depletion is channeled through the atmospheric abiotic processes. Through the atmospheric transport downward to mid- latitudes, biota elsewhere in NH becomes a subject to the energy and information transitions set by the Arctic bromine explosion.

## § 4 GAIA PARADIGM OF THE PHENOMENA
**Index: Microbial life forms, metabolic activities and metabolic excretions into environment, the sequestering of matter and the formation of physical and chemical gradients**

Tropospheric ozone amounts and concentrations of dissolved oxygen depend on surface/ ABL temperatures and sea ice melting. Surface temperatures and sea ice melting are the derivatives of the solar factor. Interdisciplinary experimental findings and observations in the Arctic extreme conditions show that variations of the solar control factor are successfully employed in the Earth's biogeochemical cycling in favor of microbial life forms. Cyanobacteria are dominated in the polar regions ([24],[26] and [3]).

Biotic processes of photosynthesis and respiration are strongly constrained by situation at the seasonal transition to the productive period. NH seasonal transition features the phenomenon of the Arctic bromine explosion. Initiated by the Arctic bromine explosion, ozone-oxygen transformation is well placed within the Living Earth context. The Living Earth (Gaia hypothesis by Lovelock and Margulis 1974; Pujol 2002) hypothesis is the theory that living organisms and inorganic material are part of a dynamic homeostatic system**.** The physical Living Earth is dominated by microbes. Among the Earth's living organisms, microbial organisms are the most special**.** We can't imagine any other life forms retaining the viability over geological times. The viability ensues from the combined effect of the single cell microbial size, population adaptability of microbes and unique features of permafrost. At the opposite end, there are the glacial erosion processes, thermal decay and solar radiation. They limit the viability of microorganisms frozen in permafrost and regulate population dynamics. The degradation processes and metabolic excretions contribute to the biotic-abiotic transformation of matter and to input fluxes into the Earth's environments, such as ocean, sea ice/ snow pack, soil ([18]) and atmosphere.
Microbial physiology is able to comply with environments lacking energy and oxygen. Instantly, at very cold Arctic temperatures, metabolic activity undergoes the temporary shutdown. At virtually no metabolism, microbiota remains viable and preserves its organization. In the adverse environment, a number of generations for microbial life are limited by longevity of the reproductive periods. The microbial life forms combine the short periods of the fast exponential reproduction and the long periods without reproduction. Paleo microbial life forms may enjoy an ultimately long life of isolated individual organism. There are cases in both polar regions when microbiota has been isolated from the cores up to 400 m deep and ground temperatures of −27°C. There are also examples of permafrost zones and biota reservoirs on asteroids in [10]. The restoration of cell metabolism and reproduction are observed, nevertheless [22].

## § 4.1. FLEXIBILITY OF ENTROPY PRODUCTION
**Index: Optimal rate of entropy production, Arctic extreme environment**

Input of external energy to the Earth's system obliges Earth's biota to produce entropy at optimal possible rate. Input of external energy to the Earth's system has never been constant or uniform. As so, many optimal production rates were exerted through the geological times. In fact, contemporary Earth's biota is represented by a variety of biota with different entropy production rates. Global vector of the Earth's development is pointed toward higher rates of entropy production and large life forms, because large organisms and organized populations yield higher rates of entropy production.

Flexibility of entropy production is particular important for the extreme environments. In case of rapid changes in input energy, microbial populations with their flexibility of entropy production, have several distinctive advantages before large organisms.    One of the grounds for the higher rates of entropy production of large organisms is a food web development. Existence of large organisms is completely relying on the food web built on the microbiota. At present, the paleo microbial life coexists with large life forms. Extreme environments hardly accommodate large organisms. Due to the specifics of entropy production in the extreme environments, the viability of microorganisms is absolutely higher than the viability of the large life forms. The Arctic environment is an example of environment that is a mostly favorable for marine microbes. It can be expected that all major surface and near-surface processes of the Earth's system in Arctic are the result of metabolic activities of the surface-dwelling microbiota.

## § 4.2. BIOTIC ORIGIN OF THE ARCTIC BROMINE EXPLOSION
**Index: Hypothesis of the biotic origin of Arctic bromine explosion, formation of seasonal NH gradients of the atmospheric species**

Arctic bromine explosion phenomena are the result of marine microbial activities. At sunlit, with erosion of permafrost, microorganisms restore physiologically. The growing microbiota population requires sufficient amounts of dissolved oxygen. Before DO will be produced by algae blooming, tropospheric oxygen is a main source for dissolved oxygen. The hypothesis of the biotic origin of Arctic bromine explosion accounts for the following findings:

(i) Bromine explosion occurs during February – May (peak March-April), with single episodes over June-August / Bromine explosion starts at coastal line and offshore locations

Abiotic processes: Initial bromine flux to atmosphere comes from eroded permafrost on coastal locations. Stormy winds and blowing snow pollute bromine species and tropospheric ozone over the Arctic sea ice/ snow pack.
Biotic processes: Bromine salt aggregates are by-product of marine microbial organisms' metabolism. The locations of aggregates' repositories are determined by the habitat of microbial organisms. At polar sunrise, microbial life is physiologically restored in Arctic SML waters. During the bromine explosion, we observe the management of SML "trade off" between the bromine- out and oxygen- in ocean fluxes.

(ii) Bromine explosion starts at sunlit (February-March)

Abiotic processes include (a) an input of stratospheric ozone to Arctic ABL, (b) activation of the bromine chemistry upon solar factor, for example, photochemical dissociation of molecular bromine. With the certain phase log, the variations in solar factor determine the timing of the first episodes of bromine explosion. Hours of sunlight and local ABL ozone concentrations prescribe the longevity of local ozone depletion. (See Figures 7-16)
Biotic processes: Ozone-oxygen conversion is required for the life sustaining microbial activities. Corresponding to the first episodes of bromine explosion and synoptic-climatic situation, intensive oxygen fluxes to SML supply the sufficient amounts of dissolved oxygen.

(iii) Bromine explosion finishes in the middle of daylight season

Biotic/ Abiotic: Net marine production regulates the dissolved oxygen amounts and provides oxygen flux to ABL. Normalization of surface ozone levels and stabilization of dissolved oxygen amounts are reached at the thresholds of the Arctic marine community biodiversity.

(iv) Bromine explosion is associated with the Arctic cold surface temperatures

Cold surface temperatures advantage several biotic-abiotic processes: Solubility of gases is higher at cold temperatures. Tropospheric oxygen fluxes to the Arctic SML/ Metabolism of Arctic microbiota is adaptable to extremely cold environments. The dissolved oxygen is consumed by marine biota. Intensification of the oxygen fluxes to the Arctic waters regenerates close to saturated DO levels and thus, supports marine life.

The order in the Earth's system involves the sequestering of matter and the formation of physical and chemical gradients by biotic and abiotic processes. The biotic bromine sequestering in the Arctic ocean precedes to the atmospheric phenomenon of bromine explosion. Bromine explosion/ tropospheric ozone depletion induces the formation of NH gradients of the atmospheric species. Through bromine explosion, polar sunrise energy is redistributed over Arctic and subarctic regions. In [8], we expressed the phenomenon formative factor by the "tropospheric constraint BrO/ Bry". For today's climatic conditions, normal March-April BrO amount is about 3e27 molecules, while the value of the phenomenal formative factor is about 0.23.

# § 5 ONE EARTH'S TYPE OF LIFE
**Index: Earth's type of life, coupled development of the Earth's life and Earths' atmosphere, "metabolic" waste control**

Due to their common origin, organic Earth's life forms share a common set of bio molecules. Biological and thermodynamic planetary development requires free energy consumption and subsequent entropy production. The Earth's life uses photon free energy to perform charge separation [9]. There is only one type of the organic life at the surface of our planet - water-like, water-solvent type of the organic life.

Coupling of the Earth's life forms and the Earth's atmosphere is built on the metabolic adaptability of life forms and the interminability of the elemental biogeochemical cycling. Prime life forms were present on the planet during low free oxygen geological periods. They should have used other than oxygen terminal oxidants for biology. Over geological times the biosphere – atmosphere complex system has been developed toward the oxygen-rich atmosphere and the advent of oxygenic photosynthesis. Modern surface dwelling life forms depend on the sufficient atmospheric $O_2$ for the synthesis of many essential biochemicals. Aerobic life forms require molecular oxygen not only as a source of free energy, but also for respiration and as protective agent against harmful UV radiation. Disposal of oxygen and reactive oxygen metabolites from the living organisms into environment is a part of the global oxygen biogeochemical cycle.

Ozone is toxic for the organic life forms. Bromine substances are also toxic for the organic life forms (IPCS Chemical Safety, report of World Health Organization on bromide ions concentrations, 1995, *http://www.inchem.org*). Ozone and bromine are highly reactive. Even so bromine substances are toxic, they are easily bio accumulated. Ozone is not assimilated, but oxygen is. Halogen chemistry avails for the metabolic waste control. Natural technologies of the metabolic waste control assume waste control not only in a body of the Earth's life, but in the environment also. In the habitat environments, ozone-oxygen conversion under bromine chemistry is one of the essential pathways of the metabolic waste control. Causality requires a temporal precedence of the sufficient amounts of halogen substances (bromine, chlorine, iodine [7]) of biotic origin to the advent of the oxygen photosynthetic production over earth surface.

The easiness of bromine bioaccumulation contains a clue regarding the role of bromine in the planet's biosphere development. We distinguish several types of life strategies ([17]) occurring in natural selection. L-selection is happening in an adverse environment. Predominantly, L-selected organisms like Bacillus circulaus are adapted to extreme environments. There are some interesting findings and models for L-selected microbial life that can be applied to the Arctic environment. Metabolic pathways for extreme environments appeared to be the most old and the most basic physiological pathways of the Earth's life. Geological records state that the Arctic is an adverse environment. Stimulated by variations in solar factor (and other outer forcing) and/or albedo, adverse abiotic loop initiate chain of the powerful reverse biotic processes. The connection between the Earth's environment and the Earth's type of life is so fundamental, that we reasonably expect that today's extreme conditions are "kept" by the paleo marine microbiota, because they support its (a) conservation and (b) metabolism.

## § 6 BROMINE METABOLISM
**Index: Terminal oxidants for biology, significance of brominated metabolites for the synthesis of essential amino acids**

Back in geological times short of free molecular $O_2$ life forms should successfully used bromine (chlorine, iodine) oxidants in their metabolism. Bromine is a powerful oxidizing chemical. It bonds with most organic compounds and metals, especially in the presence of water and especially upon illumination. It is recognized that certain bromine-related compounds are easily bio accumulating in living organisms. Not occasionally, the range of biota responsible for the production of bromine compounds is extremely wide ([15]). In [2], D. Catling compares the possible terminal oxidants for biology: F, Cl, Br, O, N, C and S. The comparison shows connection between the oxidant abundance in the Earth's system and free energy of reduction to the aqueous hydride per electron under standard conditions $\Delta G°$ (298.15 K, 1 atm), hydride bond enthalpy per electron, and the prohibitive kinetic reactivity with organics. Based on the thermodynamic factors and prohibitive kinetic reactivity, it is suggested that oxygen is superior to other terminal oxidants.

At the beginning stages of the Earth's development, the physical-chemical conditions were far away from today's standard climate conditions. In the paleo physical-chemical conditions, bromine was a demanded oxidant in the metabolic pathways of the marine life.

Many publications contain an information on the brominated substrates and on the ecological significance of bromine metabolites in biota. In his solid monograph [15], A. Neilson discloses the whole catalogue of brominated metabolites. He also provides a valuable description of halogenization biosynthesis by Hal+ reactions and Hal-reactions. The bromine compounds have a variety of roles in metabolism. The exceptionally important fact is that the brominated and chlorinated metabolites are necessary for synthesis of the essential amino acids.

Amino acids are present in all biochemical molecules. The eight amino acids are called essential, because large life forms cannot synthesize them from other compounds at the sufficient levels, so they must be obtained from the food web.

Besides that, brominated metabolites are active in the synthesis of four essential amino acids: leucine, methionine, phenylalanine and tryptophan. Clearly, those amino acids are the most ancient and developed at the times of the preoxygenic atmosphere. Paleo microbiota uses bromine compounds as metabolic agents in building amino acids and for protection against UV radiation. Inevitably, fixed by microbes, the metabolic by-products such as bromine salt aggregates must be either way introduced to the biogeochemical cycling. Introduction occurs through the biotic communities' reservoirs and/ or through degradation of these communities. Arctic region is a reservoir of the viable paleo marine microbiota, and Arctic is the arena of the ABL bromine explosion. Bromine input to the Arctic ABL is a manifestation of the biological control over bromine and oxygen cycles. Qualitative analysis of phenomenon and its consequences for the environment is available throughout the scope of atmospheric and ocean findings.

# § 7 PRO-LIFE EFFECT OF THE ARCTIC BROMINE EMISSIONS
**Index: Habitat of the Arctic ocean vs. dissolved oxygen in surface mixed layer, perturbation of NH tropospheric ozone**

Blowing winds distribute initial bromine compounds and their derivatives all over the Arctic region. In [8], we suggested the universal scenario of epidemics spreads exponentially quickly out from the source. Such scenario reflects on and stresses the underlying connectivity of the surface processes. Initially, heterogeneous bromine chemistry proceeds in the snow pack. Further, through the abiotic air-surface exchange processes, bromine species are introduced to the Arctic ABL. Relevant to exchanges with atmosphere, physical properties of snow are discussed in [4]. At present, as we ignore bromine production by aerosols, bromine family's composition and partitioning of the Arctic troposphere is determined by (a) catalytic bromine chemistry in gaseous phase (b) bromine recycling on aerosols, (c) loss due to the deposition and long-distant atmospheric transport.

Detailed review of the trace chemicals and the corresponding homogeneous and heterogeneous reactions may be found in [22]. However, mechanism and parameters of the arbitrary microbial absorption of bromine from marine environment or bromine by-product concentrating are still left to be dealt with. Sea-(sea ice/ snow pack)-air fluxes of bromine substances are rationalized in [8]. In autumn 1995, some attempts to evaluate SML bromine concentrating and sea-air fluxes were made during the Greenland expedition of ARKTIS-XI/2 of RV "Polarstern". To the best of our knowledge, all available investigations on the spring sea-air fluxes of bromine substances had been conducted at conditions where fluxes were primarily related to the bromine production and/or recycling on aerosols. Bromine salts aggregates are released to the Arctic waters at the habitat locations of microbial community. Habitat locations include the Arctic shallow waters along the shore line. Processes of the aquatic bromine release by marine microbiota are similar to the processes of nitrogen release to soils by soil bacteria.

In both cases, biogeochemical cycling prompts the species flow into the atmosphere. At polar sunrise, when bromine is emitted to the Arctic atmosphere, we observe the tropospheric ozone depletion. Excess of tropospheric ozone is converted into the molecular oxygen. The part of the molecular oxygen sinks in the Arctic waters and stimulates the springtime marine production. The dissolution of oxygen in fresh and salt water is an exothermic process. At the low Arctic ocean temperatures and high oxygen fluxes, the dissolution shifts towards the saturated levels of dissolved oxygen about 15mg/L. At the initial stage of life cycle, when the microbial activity just has been restored, organisms may consume all DO out of the water, but continuing sink of atmospheric oxygen balances depleted DO. As spring proceeds, the photosynthetic oxygen production by marine life stabilizes the dissolved oxygen levels. The levels of dissolved oxygen in SML are controlled by the water temperature and thresholds of marine life biodiversity.

At polar sunrise, Arctic bromine explosion leads to the complete depletion of the local elevations of surface ozone and the intensification of oxygen fluxes. By the major north-to-south routes, bromine- polluted polar air masses transfer to the middle latitudes. Along the transport, they perturb surface ozone of NH. Spatial - temporal variations of atmospheric bromine compounds and destruction of surface ozone signify the beginning of seasonal transition to the productive period (of oxygen photosynthetic microbiota and its food webs).

# § 8 THE FOUNDATIONAL PRINCIPLE OF MULTIPLE UNITY
**Index: Earth's multiple unity and divergence, dialectic of Earth's operations, atmosphere as an indicator to the changes in the microbiota activities, explicit introduction of biotic factor into atmospheric modeling**

At polar sunrise, the Arctic microbial organisms create conditions which are hospitable for the other life forms, for the Earth's biodiversity. Resuming the description of phenomenological model, we must conclude that biotic-abiotic unity is a goal and a driving force for existence of the non equilibrium, self regulating Earth's system. We apparently perceive a physical reality that works out on the principle of multiple unity. Credibility of the conceptual and mathematical modeling of the Earth's system and its compartments is determined by whether logic of the interminable Earth's biotic-abiotic functioning is included. The introduction of the multiple unity into the mathematical apparatus of computational models can be done through the set of the essential synchronization factors between biotic and abiotic operations.

Formulation of the possible algorithmic expressions for the interminable multiple unity demands a serious mathematical research and an advantageous knowledge of (nature) foundations. A working implementation for the principle of multiple unity/ diversity is presented in DaisyWorld models. DaisyWorld models reflect on a simplified divergence of priorities between species (black and white daisies in the original model by Watson and Lovelock). Divergence of priorities is followed from the differences in the rates of entropy production (optimal growth temperature). Divergence allows a self - regulation of DaisyWorld upon a solar radiation and a weighted albedo. Solar factor may be kept constant, but the weighted DaisyWorld's albedo is changing. As a result of the life forms' cooperation, DaisyWorld is compromised on the environment optimum for all. The differential equations of DaisyWorld couple biotic and abiotic parameters such as production and death rates, spreading threshold, albedos and optimal growth temperatures.

Modeling of the Earth's system is much more complicated. It requires a prior knowledge of the Earth's major biota groups functioning. Earth's major biota contains several mirobiota groups, which have different population dynamics, optimal temperature ranges etc. Today's state of the Earth's system reflects on oxygen as a terminal oxidant of biology. Paleo microbiota, capable to use several terminal oxidants (alternative switching between terminal oxidants), is found in a base of a food web. The estimation of climate variability completely depends on our awareness of the paleo microbiota operations.

Bromine concentrating by marine microbiota and nitrogen fixation by soil bacteria are examples of the such global scale operations. They should be the first ones introduced to the Earth's system modeling. Atmospheric modeling can be really the efficient means of understanding of the functioning of the Earth's life.
Since (a) atmosphere is a very exact indicator to the changes in the microbiota activities, and (b) we already possess an advanced, e.g. satellite, technology providing us with the consistent atmospheric data, atmospheric modeling can serve a start point in the statistical modeling of the biotic Earth ( see part 4of this study).

# § 9 SOLAR CYCLES
**Index: External energy forcing, solar cycles, microbiotic responses to the solar variability, correlation between atmospheric ozone and solar factor**

The Earth's system operations are powered by the external energy forcing. At most, consistence of the biotic-abiotic operations is handled through the integral external constraint of solar factor. Nearly all of the energy on Earth comes from Sun. Solar variability is a main element of external forcing variability ([6]). Biotic-abiotic operations correlate with solar factor, seasonally and periodically (Milankovitch's theory of insolation, Mathematical Climatology and the Astronomical Theory of Climatic Changes). For our geological period, interannual solar variability is characterized by periodicities of Hale 11 yr, Schwabe 22 yr, Gleisberg 88 yr, Suess 208 yr, 550 yr etc known as Solar Cycles. In a raw approximation, Sun is a self sustained system with an internal energy source which works as a relaxation oscillator (E. Parker, A. Pikovsky). Within each Solar Cycle, there are periods of the high and low activity of the solar dynamo. The series of solar cycles are so predictable that they can be used to forecast the next series of the climate phenomena many decades in advance. The seasonality of the incident radiation is mainly prescribed by the specific Earth's orbiting around the Sun.

At polar sunrise, we observe overlap of the several forces, such as solar factor, sea ice/ snow pack coverage, snow - atmosphere exchange, the highly dissipative atmospheric dynamics and temperature inversion of ABL. The mentioned Earth's abiotic forces depend on (a) the instantaneous solar factor (with its main component – the solar radiation) and (b) the preceding in time interannual solar variability.

A major drawback is a lack of understanding how microbiota responses to the solar variability within a solar cycle. Productive period of the polar Arctic microbiota is not continuous around the year; it begins with the bromine explosion season. During the bromine explosion season, ozone-oxygen conversion indicates on an excessive ozone flow. Excessive ozone flow has been regulated by the concurrent atmospheric circulation and preceding in time, tropic photosynthetic oxygen production. We assume that sensitivity of ozone flow to the polar sunrise ABL and sensitivity of the Arctic total column ozone to solar factor are correlated.

## § 9.1. SOLAR SYNCHRONIZATION OF ARCTIC PHENOMENA
**Index: Arctic bromine explosion at solar minima and solar maxima**

To broad out our understanding of associated phenomena, we review GOME VC BrO interannual time series which project on the bromine input to ABL (see Figure 1). From the solar minima of 1996 to the solar maxima of 2001, the differences between GOME BrO interannual time series are not so much in the total amounts BrO, but in the timing and strength of the bromine explosion episodes, and in the extent of BrO cloud. This evidence allows proposing that the total bromine input to atmosphere is less sensitive to the solar variability than ozone flow to ABL is.

During 1996–2005, the spring maxima of sea ice extent didn't change much (there is a lot of divergence in the published sea ice values). Extent of the bromine pollution of sea ice/ snow pack was compatible with the Arctic sea ice coverage. One can easily imagine conditions when total amount of emitted bromine is changing in response to the solar variability. However, consistency in the interannual March maxima of the sea ice extent doesn't let us to resolve on the possible complex dependence of the total amounts of emitted bromine from the solar factor.

For the prolonged solar minima conditions, we see that the potential of the Arctic bromine aggregates' reservoirs is sufficient enough to deplete the increased ozone flows. Applying the principle of multiple unity for the solar minima periods, we find out that Arctic biotic bromine explosion contributes to an intensification of the biogeochemical cycles. This contribution favors the extra- Arctic biota when bromine- rich air masses reach the NH mid- latitudes and cause to perturbations of the mid-latitudes ozone field and redistributions of the local energy fluxes.

For solar minima and other prolonged periods of (a) decrease in incoming solar radiation and (b) decrease in the polar total column ozone (see Table 1), role of the Arctic biota and biotic reservoirs in regulation of the Earth's biogeochemical cycles and sustainability of the Earth's system has arisen.

Non linear correlation between concurrent precursors of the Arctic bromine explosion leaves us with more questions than the answers. Resolution of the non linearity seems be possible within the phenomenological model of synchronization upon solar factor. The best available phenomenological indicators for the phenomenon synchronization are (a) timing of the first episodes, (b) temporal extent of the bromine explosion season and (c) geographical extent of BrO cloud, combined together.

Arctic springtime troposphere is a sink for stratospheric ozone. As known, stratospheric ozone layer shields our very bright planet from UV radiation. Protective stratospheric ozone layer is instantly regenerated by splitting of $O_2$ molecules by ultraviolet light. From the solar minimum to the solar maximum of Solar Cycle, the amount of UV varies as much as 400 %. During the solar minimum, the decrease in ultraviolet light received from the Sun leads to a decrease in the concentration of ozone.

## § 9.2. SEVERAL ASPECTS OF SOLAR INFLUENCE ON THE BIOTA
**Index: Compliance and resistance to the solar external forcing**

Arctic bromine explosion is a result of ocean biotic emissions. In general, rate of the (matter and energy) ocean emissions is changing cyclically on multi decadal scales which are longer than Hale 11-year (9-13 years) period of Solar Cycle. Over the short interannual period, changes in the rate of ocean emissions upon solar factor variability are smoothed for the most ocean emissions. Arctic emissions are exceptional in means that they precede to the NH transition to the production season. In spite of the fact that our understanding of a mechanism that regulates bromine emissions and their rate is approximate, we may suggest that this mechanism predefines intensity of the variety of ocean and terrestrial biotic-abiotic emissions on the global NH scale.

Arctic bromine emissions have a very straightforward effect on the tropospheric ozone field of NH. Essence of Arctic ODEs is in the ozone- oxygen transformation. In this context, local ODEs data and the global Arctic region GOME data on BrO cloud can furnish us with the necessary quantitative information about related biotic response to the solar variability.
At solar minima, biotic-abiotic response to the solar variability is as follows:

(i) According NASA GSFC (P. Newman), in 1996-1997 annual Arctic stratospheric temperatures were at low and the total column ozone depleted to about 360 DU (comparing to 450 DU in 1999). Arctic total column ozone reflects mostly on the stratospheric ozone. We assume that at polar sunrise 1996, 1997 increased UVB radiation was allowed to penetrate to the Earth's troposphere. Entered into troposphere, the amount of UVB light could increased in hundreds percent. Variations of solar activity have a radical influence on the Earth's biota. Life forms are sensitive to the changes in UVB radiation. As a result of the increased UVB, the marine microbial production in Arctic waters is projected to slow down.

(ii) The other factor of influence is availability of $^{14}C$. Solar Cycles dictate anomalies in $^{14}C$ production. Findings on $^{14}C/^{12}C$ ratio fluctuations are analyzed in ([5], [11], [23], [13]). The changes in $^{14}C/^{12}C$ ratio have an immediately impact on the microbial population dynamics. We await that sensitivity of the microbial population dynamics to the solar variability can be followed simultaneously through (a) the case of Arctic bromine explosion and (b) records on $^{14}C/^{12}C$ fluctuations.

(iii) The bromine explosion/ ozone-oxygen conversion is associated with a whole complex of the simultaneous biotic – abiotic events in the Arctic microbiota habitat. At coastal locations, the first ODEs by bromine chemistry point on the concurrent microbiota physiological restoration. During the solar minima periods, microbiota physiological restoration delays in time and production rates slow down.

(iv) The larger sea ice coverage and time delays in melting of the sea ice, guarantee that more bromine is introduced to troposphere that means intensification of ozone-oxygen transformation and larger geographical extent of BrO cloud. In Arctic waters, the metabolic bromine concentrating might be also aimed on protection of the spores' membranes from ozone. If so, the additional evidence is needed to decide about trends in the bromine input to ABL for the several consecutive years of the productive seasons at solar minima. (2007-2009 represents a good example of such solar minima period).

## § 10 INTERANNUAL VARIABILITY OF THE ARCTIC BROMINE PHENOMENA UPON SOLAR FACTOR IN 1996-2001

Life possesses the ability to resist to the external forcing. Present climate may be considered as a temporary atmosphere-biosphere attractor. In this sense, we anticipate that during the cooling periods, bromine fluxes to atmosphere would invest to the atmospheric energy budgets toward surface temperatures preceding to the cooling period. In given above perspective, at solar minima, we expect that bromine explosion would be stronger and its temporal-geographical impact would increase. Separability of episodes may be better; timing of the episodes may be more dependent on their locations.

In our study, we consider a half term of the short periodicity of 11 years. Records of sunspot counts and other proxies of solar activity are going back about 6,000 years.
They validate intuitive understanding that low sunspot counts (less than 50 sunspots) lead to a colder climate at global scale.
For the solar minima conditions, there is a strong correlation between the solar factor and stratospheric geopotential heights and temperatures (Labitzke; From the stratosphere-troposphere interactions, it also follows that at solar minima Arctic lower troposphere causes to cooler stratosphere).

## § 10.1. PREDICTABLE CHARACTERISTICS OF BROMINE EXPLOSION
**Index: Extent, timing and separability of ODEs**

The Solar Cycle 24 started in April - November 2008. NASA prediction for Solar Cycle 25 is 50 or lower counts. SSRC scientists forecast that Solar Cycle 25 could begin within first 3 years of Solar Cycle 24, which means lasting cold temperatures. A good starting point for analysis is the observational data on the former periods of minima of solar activity. We provide our results for the extent and timing of the bromine explosion phenomenon back to the previous solar minima of 1996.

Referring to the Arctic bromine explosion, we attempt to show synchronization of the Earth's system's operations upon the external solar factor. Upon the same external forcing, such as heat transfer, biotic processes have a faster response than abiotic processes. We speculate that (a) biotic bromine explosion has faster response to the solar variability than responses of the abiotic factor of sea ice coverage and (b) this is a reverse response. As a matter of fact, the speculation is based on the data on (a) GOME BrO cloud (Figures 1 -6) and (b) the Arctic sea ice coverage for 1996-2001 (NASA). For 1996-2001, GOME data shows that total amounts of the Arctic BrO didn't differ much. At the same time, seasonal and interannual extent of the sea ice has been assessed as beyond the normal. Since 2001, the extent of springtime sea ice has steadily increased. At bottom line, we expect that totals of BrO, Bry and BrO/ Bry ratio are kept in range found for 1996-2000.
For the solar minima, we expect an extension in time of bromine explosion season, stronger and better separable explosion peaks. In 2 years after solar minima (ozone quasi biennale oscillation) we expect a weaker phenomenon with the earlier first episodes.

Increase in the harmful UV radiation corresponds to (a) the seasonal springtime transition and to (b) the interannual solar minima. Each year, at sunlit**,** Arctic region has a dramatic increase in UV. Total column ozone may become slightly lower, but due to the ozone supply from stratosphere, winter and early spring ozone concentrations in the Arctic ABL are high. Unfortunately, GEM model ozone amounts were underestimated and influenced model BrO cloud strength and distribution. In our comparison of interannual bromine explosion variability, we rely on the GOME data and on the ground observations.

The analysis of separable bromine explosion events for springtime 1996-2000 (Figures 2-6) shows that for the higher sunspot counts we have more episodes. In years with less sunspot counts, these episodes start earlier and they are much weaker. The events are recorded on later dates too. The geographical extent is much smaller (10% -1% of bromine explosion extent in April 1996) than extent of episodes for the solar minimum.

The required period of observation is at least 3-4 Solar Cycles. The half Solar Cycle period observations is not enough to draw firm conclusions, but allows to show the interesting phenomenology and correspondence between control factors of the Earth's system. Table 1 shows phenomenology of abiotic and biotic processes synchronized upon solar factor.

## § 10.2. TIME SERIES FOR TOTAL AMOUNTS OF THE ARCTIC BRO

We assume that GOME BrO data is consistent. Model tropospheric ozone content was underestimated. Uncertainties for VC BrO were estimated as 20%-40% [22].
The demonstrated time series are brought for qualitative analysis only.

**Figure 1** *Total amounts BrO as observed by GOME satellite for March – June 1996-2000* were calculated above 40 N; the threshold for VC BrO is 1e13 mol/cm$^2$. BrO amounts are given in molecules. Time series for total amount BrO show range of the observed BrO as 0.8E27-3.2E27 molecules.

**Figure 1** *Total amounts BrO as observed by GOME satellite for March – June 1996-2000*
1996 was a year of solar minima (Solar Cycle 23). The highest amounts of vertical column BrO and the largest BrO cloud ware observed in 1996. The lower amounts of BrO were observed in 1999 and 2000. As shown, 7 days sampling allows detection of all peak events.

It remains an open question if we can possible diagnose the quasi-biennial oscillation based on analysis of so short interannual period. However, 1996 and 1998 time series of the total amount of BrO show similar behavior in the month of June.

**Table 1** *Phenomenology of the springtime transition to the NH productive season*
Analysis on the phenomenology corresponding to the interannual time series shown in Figure 1.* as April sea ice extent we used Goddard data area of the estimated sea ice extent and the extent excluding North Pole area not seen by sensors. Features of BrO cloud at solar maxima of 2001 are not presented.

**Figures 2-6** *Time series for the total area of bromine explosion above 40N in 1996 – 2000* are presented. The series were calculated for the observed GOME VC BrO. VC BrO threshold was taken as 1.e14 molecules/cm$^2$. Area was scaled by 1e7 km$^2$ (Figure 2) and 1e6 km$^2$ (Figures 3-6). Time period of analysis was 120 days, starting March 1.

**Figure 2** 1996 year: Maximum for total area for VC BrO above 1.e14 mol/cm$^2$ was 2.5e7 km$^2$. Episodes (peak events) were in end of March and first decade of April.

**Figure 3** 1997 year: Maximum for total area for VC BrO above 1.e14 mol/cm$^2$ was 3.6e6 km$^2$. All three episodes occurred in April.

**Figure 4** 1998 year: Maximum for total area for VC BrO above 1.e14 mol/cm$^2$ was 7e6 km$^2$. Several episodes were observed in March, and even in May. Two episodes occurred in the last decade of April.

**Figure 5** 1999 year: Maximum for total area for VC BrO above 1.e14 mol/cm$^2$ was 11e6 km$^2$. One episode happened in the second decade of March. Four episodes occurred after the second decade in April.

**Figure 6** 2000 year: Maximum for total area for VC BrO above 1.e14 mol/cm$^2$ was 15e6 km$^2$. All three episodes were in March and in the first decade of April.

**Figure 1** Time series of total amount BrO above 40N in 1996-2000

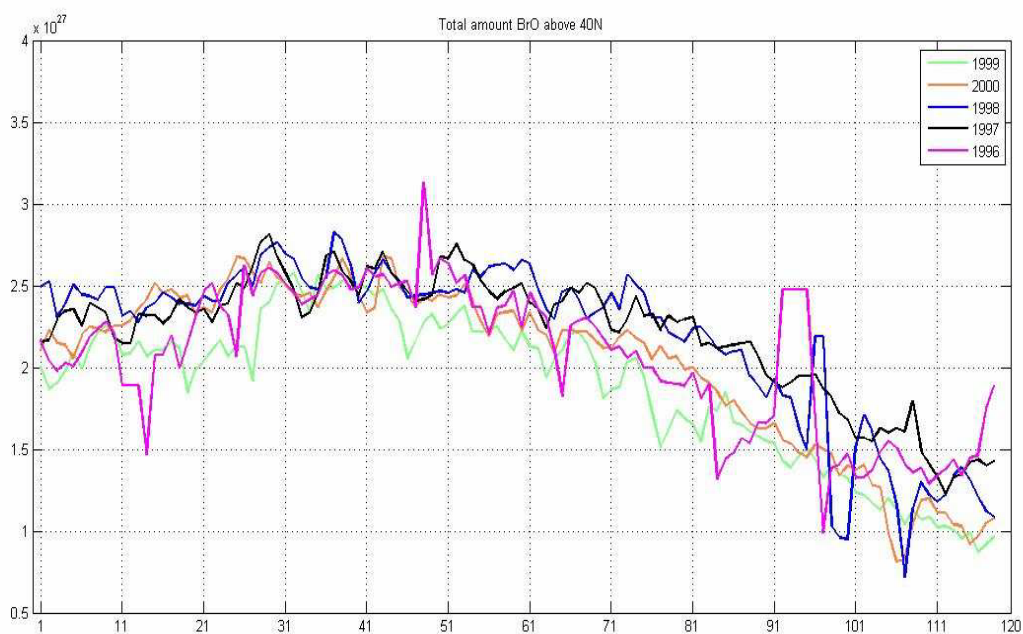

**Table 1 Phenomenology of the springtime transition to the NH productive season**

| year | Sun spots count for March | Sun spots count for April | Total column ozone DU | April sea ice extent million sq km | Extent of April BrO cloud sq km | Timing of high VC BrO | Longevity of BrO cloud above 1.5e7mol |
|---|---|---|---|---|---|---|---|
| 1996 | 9.7 | 8.4 | n/a | 14.22-12.23 | 2.5e7 | March 21 | end of June and even later |
| 1997 | 13.5 | 16.5 | 340 | 14.59-12.48 | 3.6e6 | March 27 | June 15 |
| 1998 | 53.5 | 56.6 | 420 | 14.89-12.76 | 7e6 | March 26 | June 1 |
| 1999 | 83.8 | 85.5 | 450 | 15.13-13.08 | 11e6 | March 31 | May 1 |
| 2000 | 119.9 | 120.8 | 390 | 14.63-12.51 | 15e6 | March 22 | May 5 |
| 2001 | 104.8 | 107.5 | 440 | 14.86-12.99 | Not presented | | |

Sun spots counts for March and April are smoothed monthly, data is adapted from NASA
Total column ozone is adapted from JR-25, Japan
Timing of high VC BrO starting month of March is given for column BrO above 2.5e7 mol
Longevity of BrO cloud- indicate on continuing influx of BrO into atmosphere

**Figures 2-6** Time series of total area of "bromine explosion" above 40N for visual analysis. Figure2 has a scaling different from Figures 3-6!

**Figure 2**             **Figure 3**

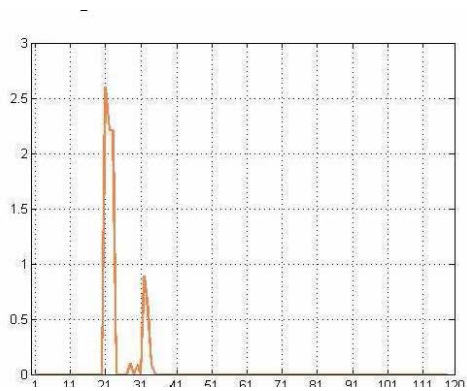 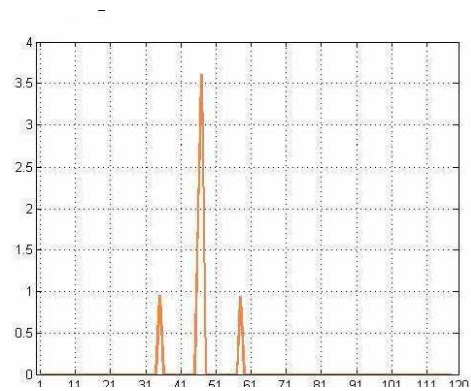

**Figure 4**             **Figure 5**

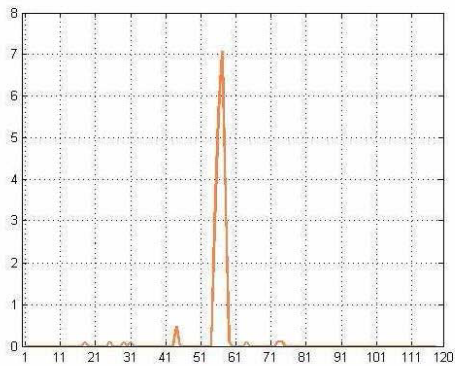 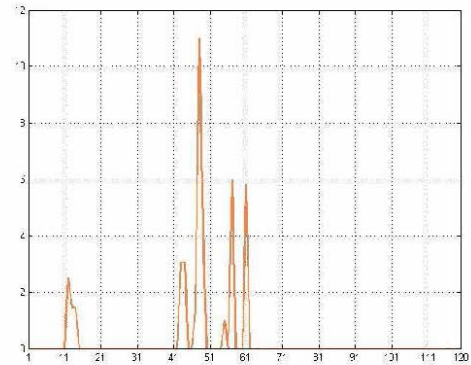

**Figure 6**

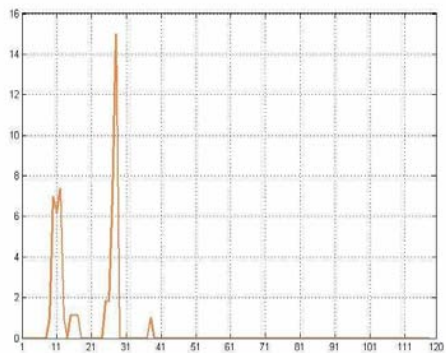

# § 11 QUALITATIVE ANALYSIS OF OZONE-OXYGEN CONVERSION
Index: Local episodes of surface ozone depletion for March-April 2001, analysis model and observed data

At this time, it is not possible to identify a particular kind of marine microbiota responsible for Arctic bromine explosion or estimate the microbial bromine inventory. Verification of the heterogeneous chemistry in sea ice/ snow pack has not been performed yet. Life times of corresponding processes are uncertain. As so, we describe the only possible qualitative analysis of ozone- oxygen conversion which is based on the modeling tropospheric data. Modeling also gives an opportunity to compare the case of the realistic bromine emissions to the reference case. Description of cases Cs1 and Cs0 is given in [8]. To show effect of the bromine explosion in NH ozone – oxygen conversion, we have chosen several Arctic (lat > $65^0$N) locations and one non Arctic coastal site of Mace Head (lat =$53^0$N). The sampling sites differ in means of closeness to a coast, in geographical latitude and longitude, elevation, environmental audit etc. We give below an analysis of timing of local ozone depletion and its longevity, which is based on calculated profiles and observations. We assume that (a) model results contain some qualitative information on time series of local ozone depletion and (b) time series of GOME VC BrO hint on the combined proceeding of the atmospheric circulation, actual bromine pollution and strength of the bromine flux to atmosphere. Comparison between Cs1 and Cs1 model surface ozone in vicinity of selected locations, allows us to conclude whether ozone was depleted by bromine chemistry.

We identify time intervals for similar local GOME (April 2001) and GEM vertical column BrO. After that, we analyze the ODEs initiated by bromine chemistry and happened at time intervals where GOME and GEM BrO columns hold the same order of magnitudes.
Ozone-oxygen conversion by bromine chemistry is calculated for March- April 2001 for four chosen locations (Table 2).

**Table 2**. *Model and observed local ODEs for April 2001*

|   | Site | Lat N | Lon -W,+E | 91-120JC sunlit hrs (1-30 Apr) | # ODE (BrO Ref) | Date of 1st ODE in Apr | Longevity 1st ODE days | Ozone loss in 1st ODE Ppb | Max ozone loss ppb/hr |
|---|---|---|---|---|---|---|---|---|---|
| 1 | Alert | 82.50 | -62.30 | 16-24 | 2 | 101 | 11d | 35 | 5 ppb/hr |
| 2 | Barrow | 71.32 | -156.61 | 14-19 | 5 | 98 | n/a | 20-15 | 5 ppb/hr |
| 3 | Summit | 72.57 | -38.48 | 14-19.5 | n/a | | | 20 | |
| 4 | Mace Head | 53.33 | -9.90 | 13-15 | n/a | | | n/a | |

Episodes of the significant depletion of local ozone for the time intervals when GOME and GEM BrO columns are compatible. n/a is written when no resemblance between model and GOME BRO is found. In order to show a long- distant impact of bromine emissions, we bring values of related model surface ozone loss and rate of loss even there is no resemblance between GEM and GOME vertical column BrO. The analysis takes in consideration the local variability of the atmospheric profiles. Local variability is analyzed within the vicinity of location of 90e3 km$^2$

SITE ALERT
Features: Arctic coastal location, the anticyclone area; long history of the observed ODEs and high GOME VC BRO

Alert is located on Ellesmere Island. It surrounds by hills and cliffs that rise at several hundred meters above the sea level. In April 2000, Alert site has experienced the prolonged ozone depletion event. Thermodynamic characteristic and ground based measurements of ozone are discussed in details in [25]. Especially interesting that observed in 2000, the dropping of ozone occurred during both dark and sunlit periods.
Summary on analysis (Figures 7-10): GEM model and GOME VC BrO show the same order of magnitude for 98-111 dates of Julian calendar (Fig.4). During this period, two model ODEs episodes Cs1 are correlated with high vertical column BrO. During depletion, model surface ozone of 40-45 ppb is depleted by rate of 5ppb/hr of sunlit. 90% of initial surface ozone is depleted by bromine chemistry.
Featured 2001 model ODE is similar to the ODE observed in April of 2000 ([25]).

For Alert, the impact of bromine chemistry on the surface ozone field is statistically significant

SITE BARROW
Features: Alaska coastal location, the anticyclone area; high GOME BRO column

On three sides surrounded by Arctic Ocean, Barrow is located in polar Alaska. Despite the extreme northern location, temperatures at Barrow are moderated by topography. Barrow and vicinity sit on tundra permafrost.
Summary on analysis (Figures 11-13): GEM model and GOME VC BrO show the same order of magnitude for 90-119 dates of Julian calendar (Fig.7). For April 2001, the remote from each other, Barrow and Alert sites experience high concentrations of BrO almost simultaneously. With some overestimation for 98-100 dates Julian calendar, model vertical column BrO fits GOME data. As shown, April model surface ozone varies between 35 and 0 ppb. Calculated ozone depletion up to 20 ppb occurs due to bromine chemistry.
Due to the different from Alert dynamic conditions, Barrow BrO column is subject to 1-2 days oscillations. Reflecting on day time and night time chemistry, model BrO also shows the distinctive diurnal variations.

For Barrow, the impact of bromine chemistry on the surface ozone field is statistically significant

SITE SUMMIT
Features: far from a coast, the cyclone area; GOME BrO column is lower than at Alert and Barrow

Summit is located in Greenland, about 360 km from the east coast and 500 km from the west coast, and 200 km from ice sheet.
Summary on analysis (Figures 14-16): No reasonable fit between GEM VC BrO and GOME data is shown. GOME BRO column lower than at Alert and Barrow but show a steady presence for surface BrO. Even the model amounts don't resemble BrO data, calculated ODEs show the impact of bromine chemistry. Ozone depletion up to 20 ppb occurs due to bromine chemistry.

Reflecting on day time and night time chemistry, model BrO also shows distinctive diurnal variations.

For Summit, statistically significance of the impact of bromine chemistry on the surface ozone field cannot be verified based on the one season data

SITE MACE HEAD
Features: coastal location in the Northern Atlantic

Mace Head is located on a western coast of Ireland. It is exposed to the air masses of Northern Atlantic.
Summary on analysis (Figures 14-16): For April 2001, maximum of GOME VC BrO is twice as low as at Alert site. GEM model BrO does not resemble GOME BrO and significantly less than GOME BrO. April model surface ozone varies between 18 and 55 ppb. Because of underestimated model BrO amounts at Mace Head, we were unable to analyze impact of bromine chemistry on surface ozone field. Model ozone levels may be also underestimated. Model surface BrO mix ratio follows very close to the model BrO column.

For Mace Head, the possible reason for a failure to represent the local GOME VC BRO is a model underestimation of the Arctic region ozone. Model VC BrO doesn't resemble GOME vertical column BrO.

For Alert, the impact of bromine chemistry on the surface ozone field is statistically significant.

**Figure 7**   Alert. Calculated surface ozone for Cs0, no bromine input to ABL

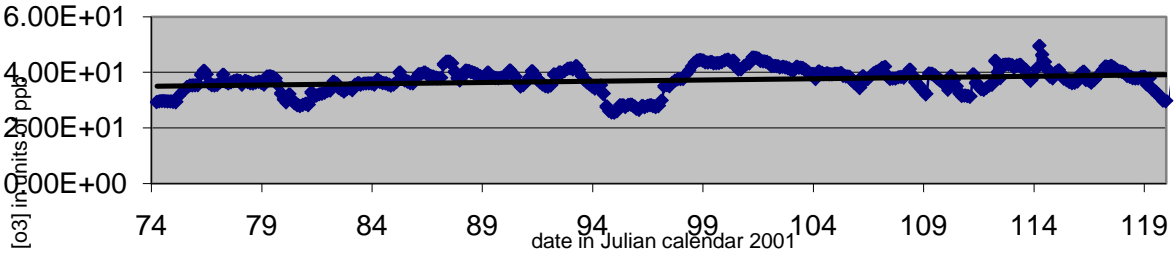

**Figure 8**   Alert. Calculated surface ozone for Cs1, bromine input to ABL

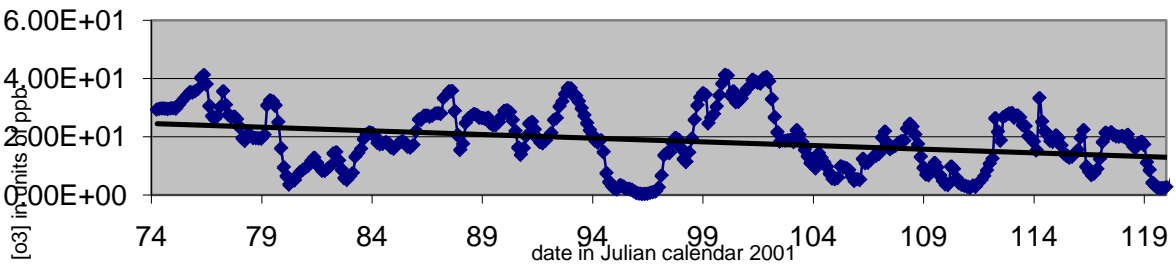

**Figure 9**   Alert and neighborhood. Calculated surface ozone for Cs1 bromine input to ABL

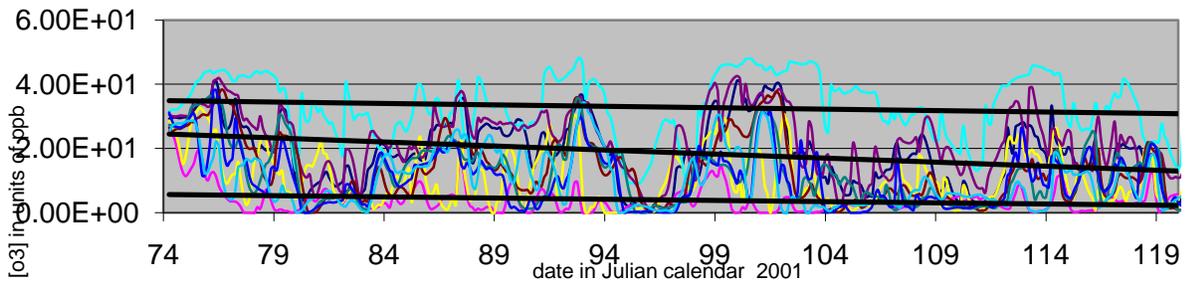

**Figure 10**   Alert.   Calculated for Cs1 and GOME VC BrO

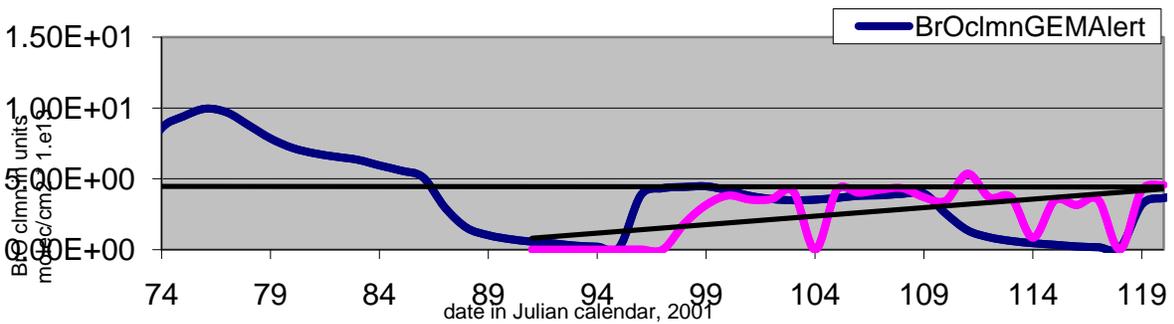

For Barrow, the impact of bromine chemistry on the surface ozone field is statistically significant.

**Figure 11**   Barrow and vicinity. Calculated surface ozone for Cs0, no bromine input to ABL

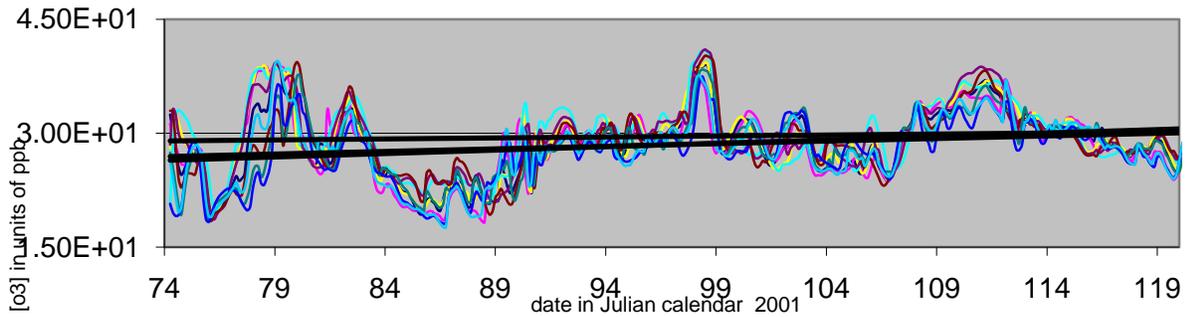

**Figure 12**   Barrow and vicinity. Calculated surface ozone for Cs1, bromine input to ABL

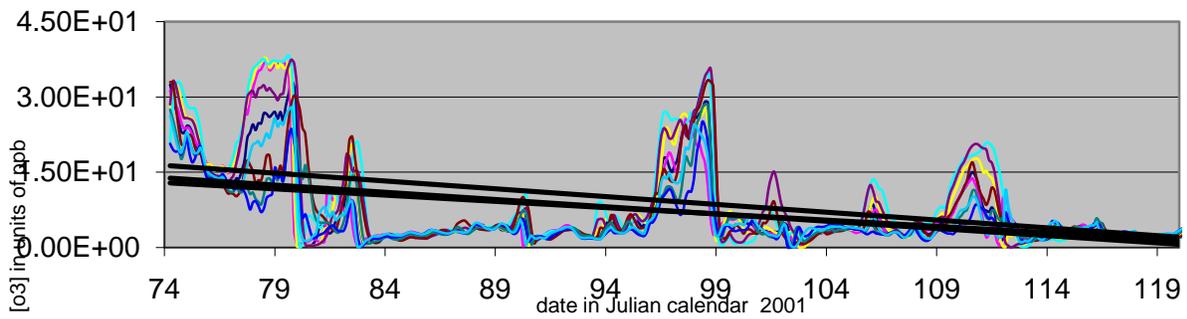

**Figure 13**   Barrow. Calculated for Cs1 and GOME VC BrO

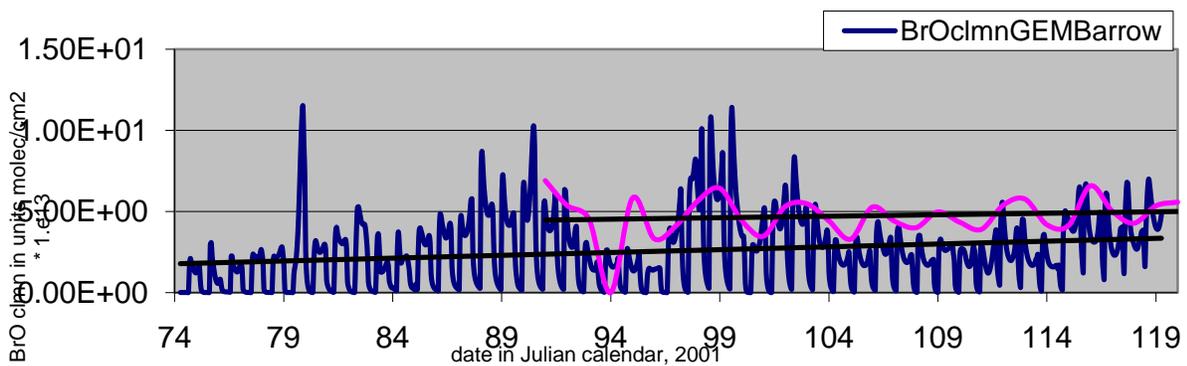

For Summit, statistically significance of the impact of bromine chemistry on the surface ozone field cannot be verified based on the one season data

**Figure 14**   Summit and vicinity. Calculated surface ozone for Cs0, no bromine input to ABL

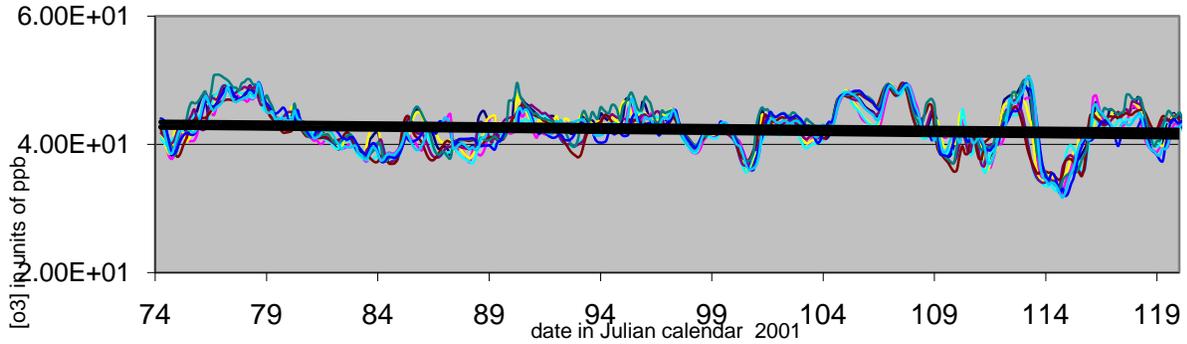

**Figure 15**   Summit. Calculated surface ozone for Cs1, bromine input to ABL

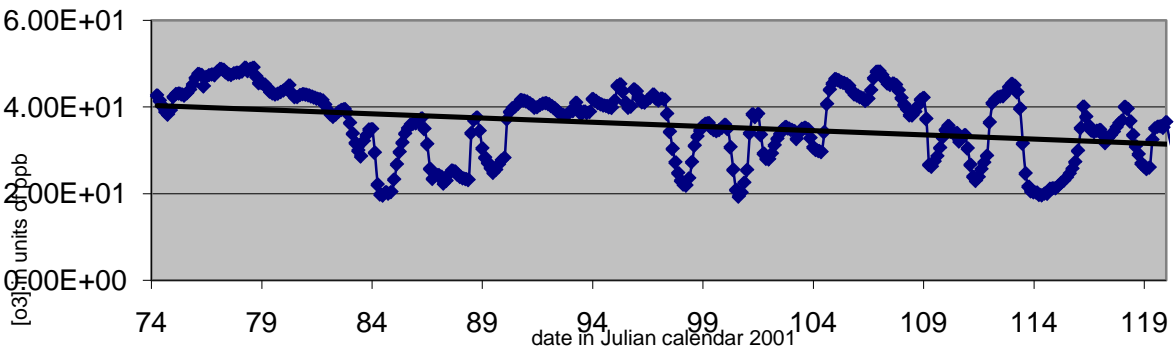

**Figure 16**   Summit. Calculated for Cs1 and GOME VC BrO

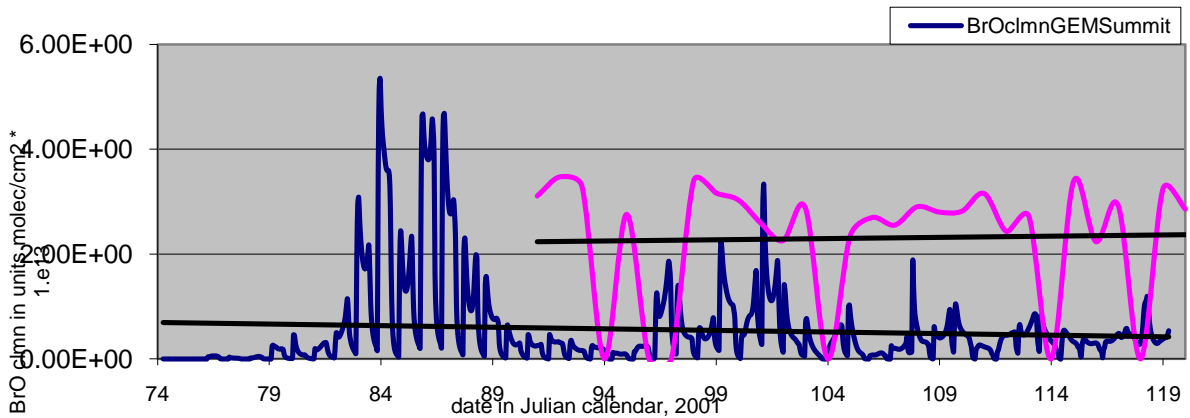

For Mace Head, the underestimated model ozone and BrO don't resemble GOME vertical column BrO.

**Figure 17**   Mace Head. Calculated surface ozone for Cs0, no bromine input to ABL

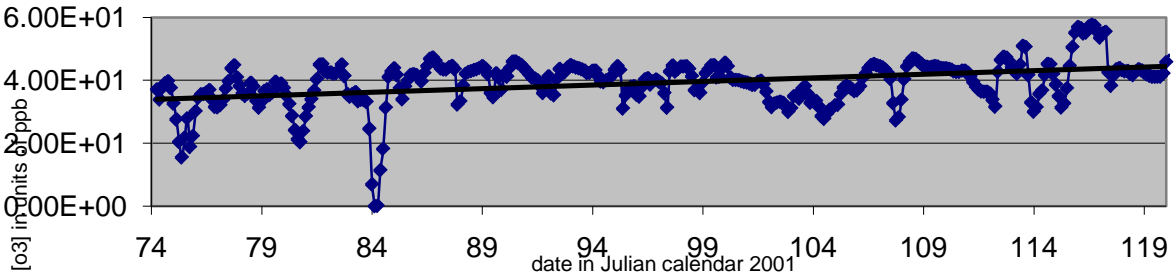

**Figure 18**   Mace Head. Calculated surface ozone for Cs1, bromine input to ABL

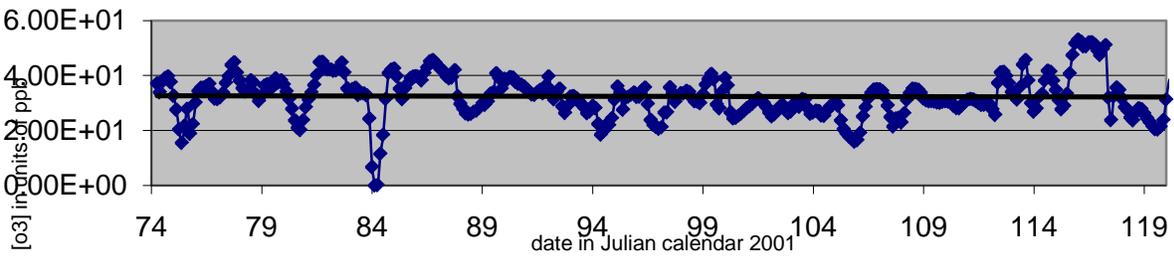

**Figure 19** Mace Head.   Calculated for Cs1 and GOME vertical column BrO

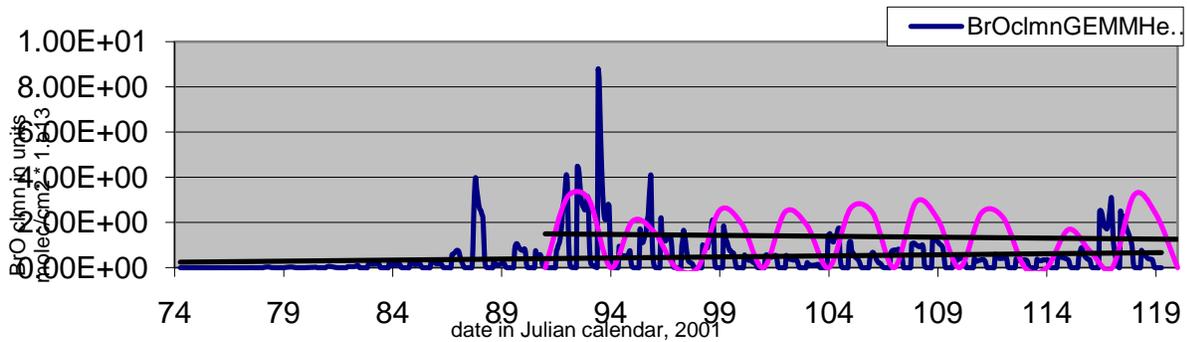

**Figure 20** Mace Head. Calculated BrO column and calculated surface BrO mix ratio for Cs1 with bromine input to ABL

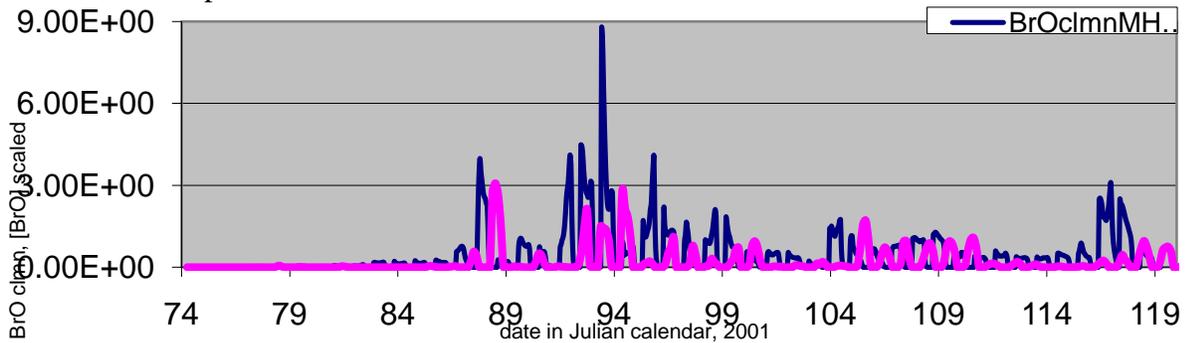

# § 12 EARTH ATMOSPHERE-CLIMATE CONTROL FACTORS
**Index: Trademark of bromine explosion, relative simplicity of understanding of the Arctic phenomena, atmosphere-climate control factors**

The importance of the establishment of biotic origin for the Arctic bromine explosion is quite apparent. Biogeochemical cycles of bromine and oxygen are interwoven through the complex of biotic-abiotic processes and synchronized upon solar factor. Seasonality and interannual periodicity of the solar factor impend seasonal and interannual patterns in biogeochemical cycles and in the earth atmosphere-climate system. Recognition of the biotic origin of the Arctic bromine explosion is a big step toward the Gaia shift in the popular and scientific perception. Bromine input to Arctic ABL leads to the catalytic depletion of surface and near-surface ozone (ozone-oxygen conversion) and to the consequent intensification of fluxes for related atmospheric substances. Owing to the recycling on aerosols and long-distance transport of the Arctic air masses, bromine compounds of polar origin are distributed toward mid latitudes of the Northern Hemisphere.

Paleo marine microbiota stands behind ocean bromine concentrating. Tropospheric bromine explosion must have been existing throughout geological times, including the prolonged periods of darkness and lack of warmth. Associated phenomena of the Arctic bromine explosion is an example of the enhanced "abiotic" ozone-oxygen transformation. Enhancement and augment of the energy-matter transformations is a way the Earth's life control factor signifies of itself and its purpose of existence.

Even though, regulation of temperature conditions is absolutely required for development of the Earth's life, climate regulation cannot be reduced to the regulation of one variable (temperature) only. The earth temperature field and infra-red radiation are coupled in a large extent with ozone-oxygen content of atmosphere**.** At low temperatures and sunlight, and at presence of heterogeneous phase, ozone-oxygen transformations upon halogen (e.g. bromine) chemistry is a basic pathway of the ozone-oxygen regulation. This pathway is realized in (1) polar stratospheric clouds (and influence polar vortex) and (2) at surface and near-surface in the polar Arctic. Evidently, bromine explosion has larger expansion at solar minima. Bromine explosion becomes very important for the climate stability and changes during the periods similar to solar minima. In the informational context, Earth's life uses the atmospheric medium as a fast communication channel: every spring, bromine-polluted air masses convey "message" to the NH biota about seasonal transition to the production phase of life cycle.

Biotic processes are chiefly responsible for maintenance of the optimal entropy state of atmosphere [14]. Biota includes several foundational groups of microbial organisms. Arctic marine microbiota is a driving force behind the Arctic ocean bromine emissions at polar sunrise. The ocean biotic emissions in the Arctic adverse environment hold a unique significance for the Earth's system operations and the Earth's ozone-oxygen conversion, because by means of the atmospheric medium they (1) activate surface ozone-oxygen conversion in polar Arctic, (2) intensify atmosphere- ocean exchange, (3) and signal and stimulate the beginning of seasonal production of NH surface-dwelling biota.
The distinctiveness and relative easiness for qualitative analysis is a trademark of bromine explosion. BrO cloud spatial- temporal extent, importance and uniqueness of phenomenon distinct it from the other biotic processes controlling the Earth's atmosphere-climate system and make the bromine explosion exceptionally valuable for the development of methodology of the atmospheric record and modeling after the Earth's system biotic processes.

Earth's microbiota (marine microbes, soil bacteria and so on) manages the redirection of energy and matter cycling. Arctic bromine explosion is a result of ocean biotic emissions into the earth atmospheric medium. The phenomena illustrate major concepts of the Gaia hypothesis.
J. Lovelock stated in the original Gaia book [12] that given the possibilities, the biosphere may multiply in the future by colonizing other planets using its Earth's biotic pumps to environment. Ocean bromine concentrating and the following bromine input to the Arctic ABL exhibit a striking similarity to the nitrogen fixation/ denitrification processes.

Bromine explosion is the instance of the basic Earth's system's biotic pumps. In the biotic-abiotic pairing, earth abiotic < chemical reactions> processes are secondary to the biotic < control over conditions for earth chemistry> processes. Owing to the existence of the biotic-abiotic pairs of reversible processes which establish the seasonal weather and stable climates, the Earth's system has recreated the thermodynamically non equilibrium atmosphere. There is a tight linkage between the Earth's life activities based on the multiple unity of the Earth's system's operations.
Biogeochemical cycling is a universal instrument providing biotic Earth's sustainability against the inconvenient loads and external forcing. Microbiotic loads (re-)create an environment optimal for the other life forms too. High levels of the near-surface and surface ozone are very inconvenient for the surface dwelling microbiota and its food webs. At polar sunrise, Arctic bromine explosion leads to the complete depletion of the surface ozone and to the consequent intensification of oxygen fluxes. Using biogeochemical principles in the atmospheric modeling, we can explore both normal and critical loads of the tropospheric bromine species of the Arctic ocean origin. Critical loads are the amounts of BrO and Bry at which climate change effects of bromine air pollutants start to occur. Normal loads are appropriate to the present-day perturbations of the ozone field. They are effectively regulated through the formative parametric constraint [8]. In the last 20-year period, normal loads of bromine compounds had appeared to match the really large variations of the observed ozone flow into the Arctic ABL. Normal loads of tropospheric amount BrO set the reversible biotic impact on the adverse environmental conditions (e.g. ozone-oxygen conversion, preventing harmful UVB or high <toxic> ozone concentrations)

Mathematical conceptualization of Br fluxes depends on (a) our understanding of a mechanism of the Arctic bromine activation and on (b) our evaluation of the regional bromine inventories. GOME satellite data and ground station measurements are used as a model reference. GOME data indirectly indicates on emissions totals and provides statistical information on the spatial and temporal separability of BrO cloud comparable to the statistics of the model case study. We also perform the qualitative interannual analysis of the time series of total amounts GOME BrO. Specific attention is paid for bromine explosion variability during prolonged periods of a solar minima. At solar minima ( e.g. 1996, strong bromine explosion), the decrease in UV received from the Sun leads to a decrease in the concentration of ozone, allowing the penetration of the increased UVB to the surface boundary layer. We show that the overall quality of the model springtime NH depressed ozone field is heavily dependent on the bromine fluxes.

## REFERENCES

_